\documentclass[amsmath,amssymb,eqsecnum,showpacs,nofootinbib,superscriptaddress,twocolumn]{revtex4-1}
\usepackage{chngcntr}

\usepackage{graphicx}
\usepackage{bm}
\usepackage{dcolumn}
\usepackage{color}

\usepackage{graphicx, tikz}% Include figure files
\usepackage{caption, subcaption}
\usepackage{amssymb,amsmath, mathrsfs}
\usepackage{amssymb, graphics, setspace}
\usepackage{hyperref}
\usepackage[all]{xy}

\newcommand{\rc}[1]{}

\newcommand{\be}{\begin{equation}}
\newcommand{\ee}{\end{equation}}
\newcommand{\bea}{\begin{eqnarray}}
\newcommand{\eea}{\end{eqnarray}}

\begin{document}

\counterwithout{equation}{section}

\author{Daniel~Pe\~{n}alver}
\email{Daniel.Penalver@ific.uv.es}
\affiliation{Departamento de F\'isica Te\'orica and IFIC, Universidad de Valencia-CSIC, Calle Dr. Moliner 50, 46100 Burjassot, Spain}
\author{Marco~De~Vito}
\email{marcodev1996@gmail.com}
\affiliation{Studio Filosofico Domenicano, affiliato alla Facolt\`a di Filosofia della Pontificia Universit\`a `San Tommaso d'Aquino' in Roma, Piazza San Domenico 13, 40124 Bologna, Italy}
\author{Roberto~Balbinot}
\email{Roberto.Balbinot@bo.infn.it}
\affiliation{Dipartimento di Fisica e Astronomia dell'Universit\`a di Bologna and INFN sezione di Bologna, Via Irnerio 46, 40126 Bologna, Italy}
\author{Alessandro~Fabbri}
\email{afabbri@ific.uv.es}
\affiliation{Departamento de F\'isica Te\'orica and IFIC, Universidad de Valencia-CSIC, Calle Dr. Moliner 50, 46100 Burjassot, Spain}
%\affiliation{Centro Studi e Ricerche E. Fermi, Piazza del Viminale 1, 00184 Roma, Italy}
%\affiliation{Dipartimento di Fisica dell'Universit\`a di Bologna and INFN sezione di Bologna, Via Irnerio 46, 40126 Bologna, Italy}
%\affiliation{Laboratoire de Physique Th\'eorique, CNRS UMR 8627, B\^at. 210, Universit\'e Paris-Sud 11, Univ. Paris-Saclay, 91405 Orsay Cedex, France}

%\author{Daniel~Pe\~{n}alver}
%\affiliation{Departamento de F\'isica Te\'orica and IFIC, Centro Mixto Universidad de Valencia-CSIC, C. Dr. Moliner 50, 46100 Burjassot, Spain.}
%\email{Daniel.Penalver@ific.uv.es}
%\author{Marco de Vito}
%\affiliation{affiliation}
%\email{marcodev1996@gmail.com}
%\author{Roberto Balbinot}
%\affiliation{Dipartimento di Fisica dell'Universit\`a di Bologna and INFN sezione di Bologna, Via Irnerio 46, 40126 Bologna, Italy,}
%\email{balbinot@bo.infn.it}
%\author{Alessandro Fabbri}
%\affiliation{Departamento de F\'isica Te\'orica and IFIC, Centro Mixto Universidad de Valencia-CSIC, C. Dr. Moliner 50, 46100 Burjassot, Spain.}
%\email{afabbri@ific.uv.es}

\title{Acoustic black holes in BECs with an extended sonic region}

\begin{abstract} 
In the context of Hawking-like radiation in sonic black holes formed by BECs we investigate the modifications of the emission spectrum caused by a finite width of the sonic transition region connecting the subsonic to supersonic flow.
\end{abstract}

\date{\today}
\maketitle

Acoustic black holes (BH) \cite{unruh} formed by Bose-Einstein condensates (BEC) undergoing transonic motion \cite{gacz1, gacz2}  have shown the presence of Hawking-like radiation \cite{hawking74, hawking75}.
The detection is however indirect, since what has been measured are the correlations between the Hawking particles and their negative energy partners \cite{bffrc, cfrbf, jeff2016, jeff2019, jeff2021}.
The transition from subsonic to supersonic flow usually occurs in an infinitely thin surface, the sonic horizon of the acoustic BH metric according
to the gravitational analogy \cite{livrevrel}.

In this letter we shall investigate the case in which this transition occurs in a layer of finite extension, i.e the flow velocity equals the speed
of sound over an extended region. We will use a simple toy model in which the condensate is step-wise homogeneous \cite{rpc, mfr, lrpc}. This kind of models are widely used in the literature and we refer the reader to the pedagogical discussion of it given in Ref. \cite{capitolo-libro}. Even if the gravitational analogy in these configurations does not hold (the surface gravity of the horizon is formally infinite), Hawking-like thermal emission is still present at low frequency with an effective temperature fixed by the healing length.

As it is well known \cite{pi-st} writing the bosonic field operator in the form $\hat\Psi(t,\vec x)=\Psi_0(\vec x)\,[1+\hat \phi(t,\vec x)]$, one has that the condensate wave-function $\Psi_0$ is governed by the Gross-Pitaevski equation
\begin{equation}\label{gp}
i\hbar\frac{\partial \Psi_0 }{\partial t} = \left(-\frac{\hbar^2}{2m}\vec \nabla^2 + V_{ext} + g |\Psi|^2 \right)\,\Psi_0
\end{equation}
and the quantum fluctuation $\hat\phi$  by Bogoliubov-de Gennes equation
\begin{equation}\label{bdg}
i\hbar \frac{ \partial \hat \phi}{dt}= - \left( \frac{\hbar^2 \vec \nabla^2}{2m} + \frac{\hbar^2}{m}\frac{\vec \nabla \Psi_0 }{\Psi_0} \vec \nabla\right)\hat\phi +ng (\hat\phi + \hat\phi^{\dagger}).
\end{equation}
$V_{ext}$ is the external potential and $g$ the atom-atom interaction coupling.
In a homogeneous (quasi) 1D condensate, the condensate wave function is 
\be \label{wfc} \Psi_0=\sqrt{n}e^{ik_0x-\omega_0t}\ , \ee
with $n$ the density,  $v_0=\frac{\hbar k_0}{m}$ the flow velocity and $\hbar \omega_0=\hbar^2 k_0^2/ (2m)+gn+V_{ext}$.
Exploiting the stationarity of the configuration we write the fluctuation field $\hat \phi$  in the form 
\begin{equation}\label{frep}
\hat\phi (t,x) =\sum_j  \left[ \hat a_j \phi_j (t,x) + \hat a_j^{\dagger} \varphi_j^*(t,x) \right]\ ,\end{equation}
where the modes $\phi,\varphi$ take the plane-wave form 
 \begin{equation}\label{mdf}
\phi_{\omega}=D(\omega)e^{-i\omega t+ik(\omega)x},\  \varphi_{\omega}=E(\omega)e^{-i\omega t+ik(\omega)x}\ .
\end{equation}
$\omega$ and $k$ satisfy the dispersion relation 
\begin{equation}\label{nrela}
(\omega-v_0k)^2=c^2\left(k^2+ \frac{\xi^2 k^4}{4}\right)\ ,
\end{equation}
with $c=\sqrt{\frac{gn}{m}}$ the speed of sound and $\xi=\frac{\hbar}{mc}$ the healing length,
and the normalization factors $D,E$ take the form
\be\label{D}
D(\omega) =  \frac{\omega -v_0 k+\frac{c\xi k^2}{2}}{\sqrt{4\pi \hbar n c\xi k^2\left| (\omega-v_0k) \left(\frac{dk}{d\omega}\right)^{-1} \right| }},\ee
\be\label{E}
E(\omega) = -\frac{\omega -v_0 k-\frac{c\xi k^2}{2}}{\sqrt{4\pi \hbar nc\xi k^2\left| (\omega-v_0k) \left(\frac{dk}{d\omega}\right)^{-1} \right| }}\ .
\ee
The sign of the quantity $\omega-v_0k=\pm c\sqrt{k^2+\frac{\xi^2k^4}{4}}$ refers to the positive and negative branch solution of (\ref{nrela}) and gives the norm of the modes. 
At fixed $\omega$, Eq. (\ref{nrela}) is a fourth order equation in $k$. It admits four solutions $k_i(\omega)$ and in general $\phi$ can be expressed as a linear combination of four plane waves constructed with the corresponding $k_i(\omega)$ as follows:
\be \label{exp-phi}
\phi_\omega(x,t)=e^{-i\omega t}\sum_{i=1}^4 A_i(\omega)D_i(\omega)e^{ik_i(\omega)x}\ , \ee
%inserire eq 1.35 del nostro paper " Understanding... " 
where $A_i$ are the amplitudes of the modes. A similar expansion holds for $\varphi$.
Before introducing our model we recall the basic features of the solutions of the dispersion relation. 

\begin{figure}[h]
\centering
\begin{subfigure}[h]{0.5\textwidth}
{\includegraphics[width=2.5in]{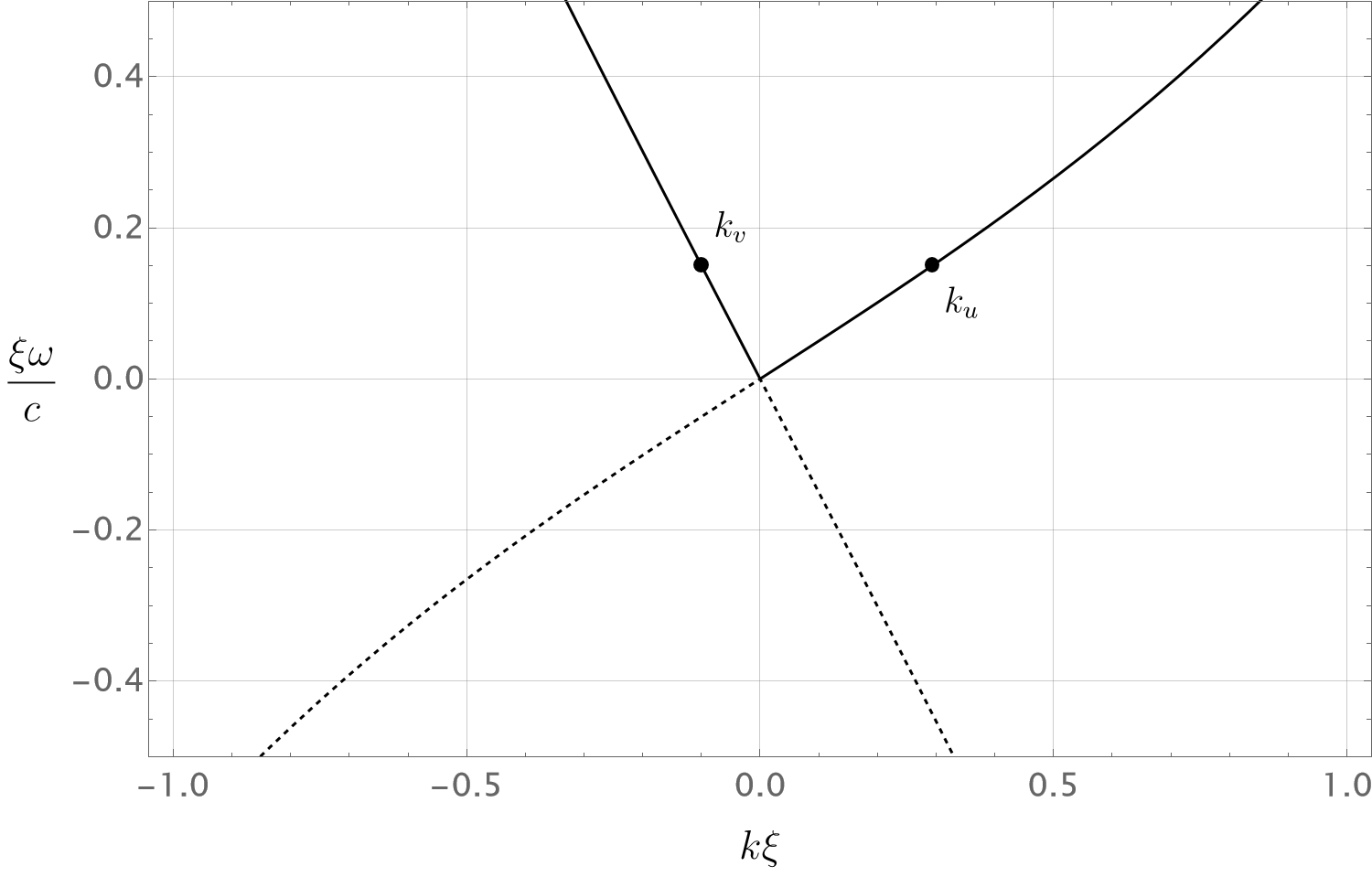}}
\caption{Subsonic dispersion relation ($\frac{|v_0|}{c}=\frac{1}{2}$).}
\label{1a}
\end{subfigure}
\begin{subfigure}[h]{0.5\textwidth}
{ \includegraphics[width=2.5in]{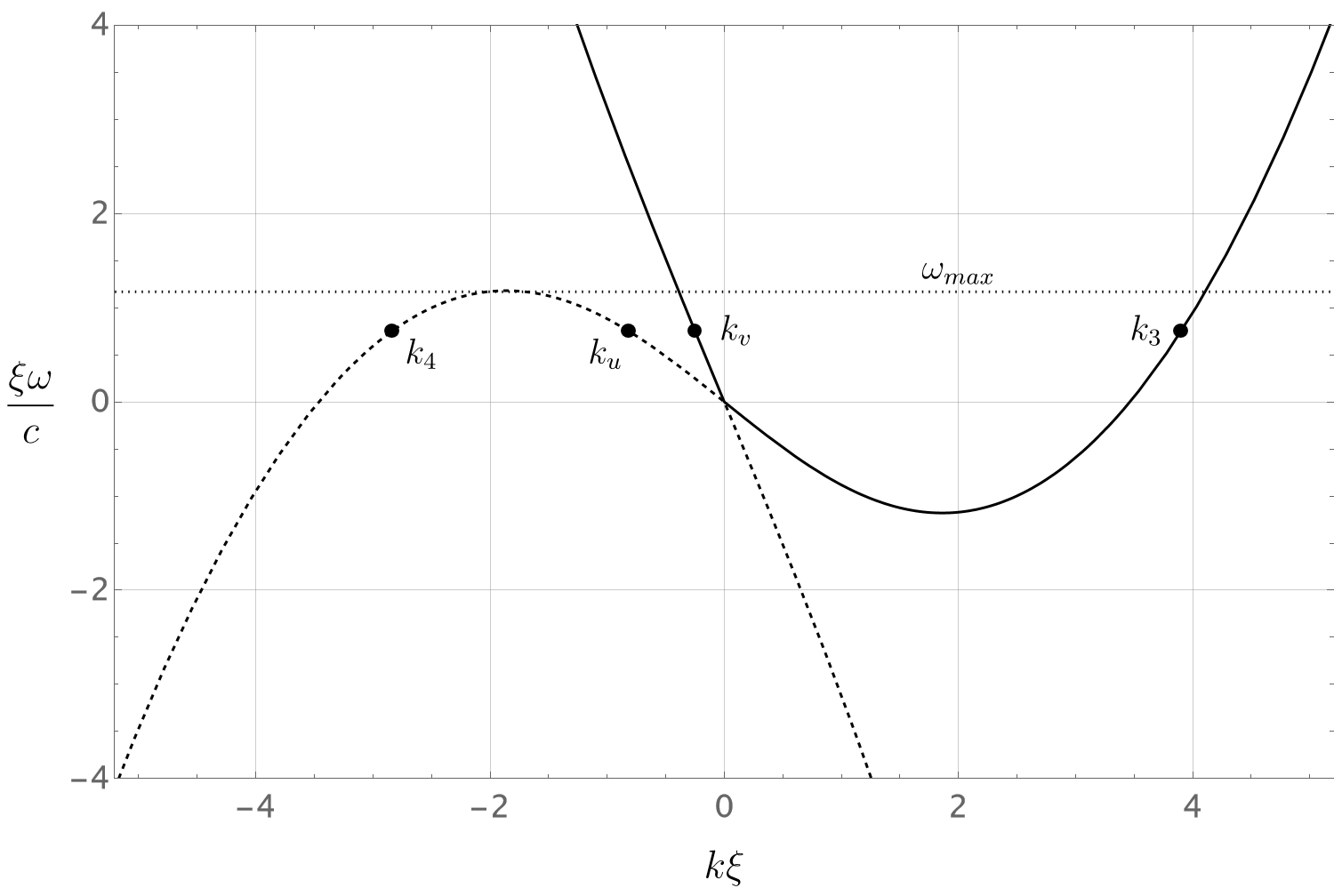}}
\caption{Supersonic dispersion relation ($\frac{|v_0|}{c}=2$).}
\label{1b}
\end{subfigure}
\begin{subfigure}[h]{0.5\textwidth}
{\includegraphics[width=2.5in]{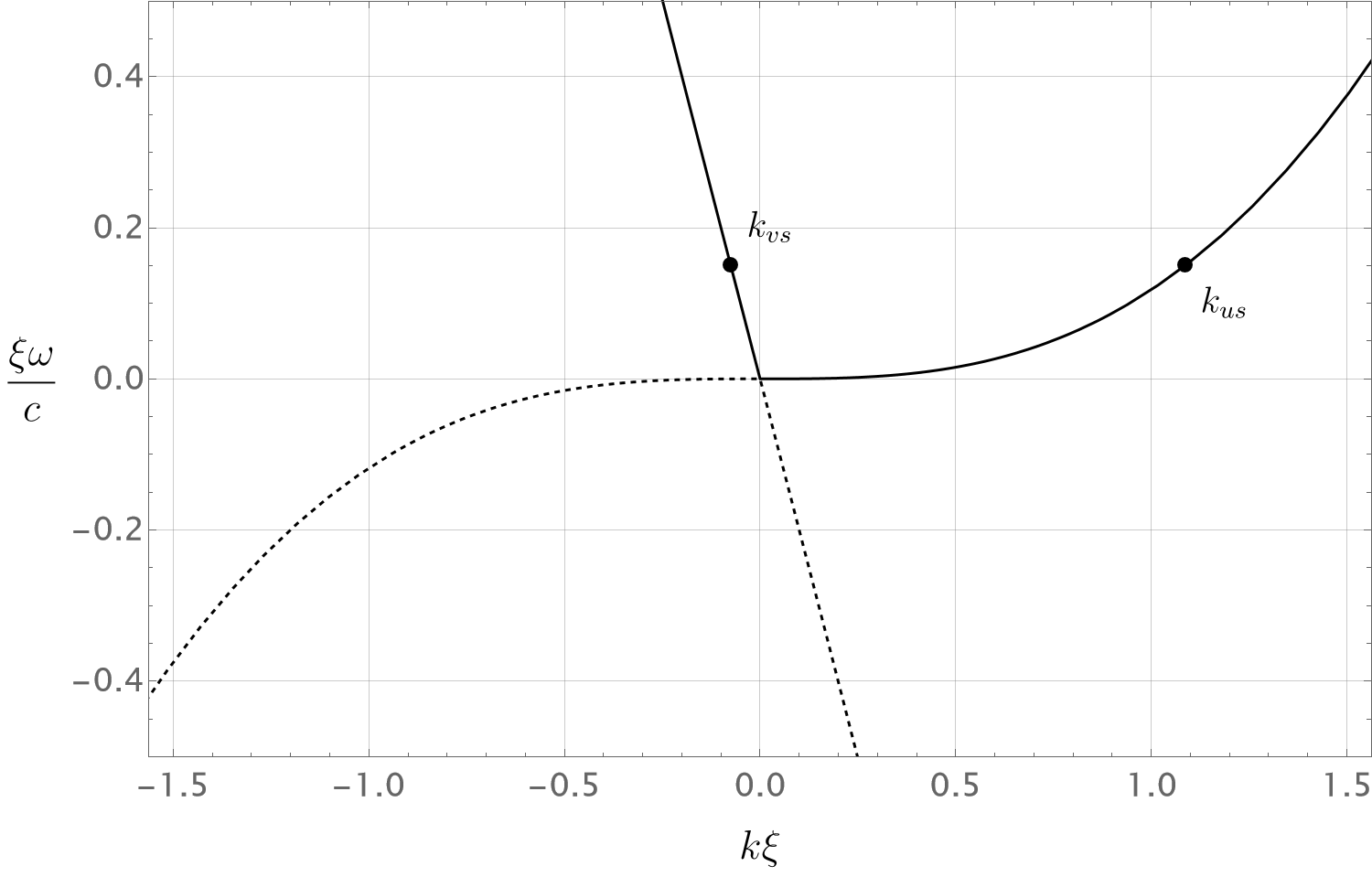}}
\caption{Sonic dispersion relation ($\frac{|v_0|}{c}=1$).}
\label{1c}
\end{subfigure}
\caption{\label{fig1} Plot of the real solutions to the dispersion relation (\ref{nrela}) 
%(here and in all the plots that follow $\frac{\hbar}{m}=1$), 
in the subsonic (a), supersonic (b) and sonic (c) cases. Solid/dotted lines represent the positive/negative branches.}
\end{figure}

For a subsonic homogeneous condensate, $|v_0|<c$ (we consider $v_0<0$, i.e. the flow is leftward), of the four solutions $k_i(\omega)$ to the quartic equation (\ref{nrela}) two are real and two are complex conjugates. The real solutions are displayed in Fig. (\ref{1a}), and correspond to a right moving wave $k_u(\omega)>0$ (in the hydrodynamic limit, $\xi\to 0$,  $k_u= \frac{\omega}{c+v_0}$) and a left moving wave $k_v(\omega)<0$ (in the hydrodynamic limit $k_v= \frac{\omega}{v_0-c}$).  The two complex conjugate solutions (we call them $k_\pm(\omega)$, $\pm$ refers to the sign of the imaginary part)  are completely dispersive (i.e. they disappear in the hydrodynamic limit) and explode either at $x=+\infty$ or at $x=-\infty$.

%\begin{figure}
%\includegraphics[width=0.8\columnwidth]{rel-disp-sub.jpg} 
%\caption{Subsonic dispersion relation.
 %\label{funo}}
%\end{figure}

For a supersonic homogeneous condensate ($|v_0|>c$) and as long as $0<\omega<\omega_{max}\sim \frac{1}{\xi}$ \cite{mapa} eq. (\ref{nrela}) admits, instead, $4$ real solutions: two left moving `hydrodynamical' $k_v, k_u$ modes (which survive in the hydrodynamical limit) and two `dispersive' right-moving ones $k_3,k_4$. They are displayed in Fig. (\ref{1b}). Note that $k_3,k_4$ are propagating upstream (their group velocity $\frac{d\omega}{dk}$ is positive) in spite of the supersonic character of the underlying flow. In addition to the number of real solutions, the novelty with respect to the subsonic case is that the $k_u, k_4$ modes belong to the negative-branch solution of the dispersion relation and, thus, are associated to negative norm modes. \footnote{Above $\omega_{max}$ the  negative energy branch disappears and we have, as in the subsonic case, two positive norm modes, one right moving and the other left moving, and two complex conjugate solutions $k_\pm(\omega)$.} 

%\begin{figure}
%\includegraphics[width=0.8\columnwidth]{rel-disp-super.jpg} 
%\caption{Supersonic dispersion relation.
 %\label{fdue}}
%\end{figure}

When the condensate velocity equals $c$, $|v_0|=c$, the dispersion relation (\ref{nrela}) admits two real solutions $k_{us}, k_{vs}$ (depicted in Fig. (\ref{1c})), associated to positive norm right moving ($k_{us}$) and positive norm left moving ($k_{vs}$) modes,  %$k_{us}, k_{vs}$
and two complex conjugate ones $k_{+s}, k_{-s}$. % corresponding to decaying and growing modes. 
%\begin{figure}
%\includegraphics[width=0.8\columnwidth]{rel-disp-sonic.jpg} 
%\caption{Sonic dispersion relation.
 %\label{fcinque}}
%\end{figure}
Thus qualitatively the sonic dispersion relation is similar to the subsonic one. The important quantitative difference regards $k_{us}$: while $k_{vs}$ is the $|v_0|\to c$ limit of the subsonic one (at small frequency $k_{vs}\sim -\frac{\omega}{2c}$), $k_{us}$ is completely dispersive ($k_{us}\sim \frac{2}{\xi}(\frac{\xi \omega}{c})^{1/3}$).

\begin{figure}
\includegraphics[width=0.9\columnwidth]{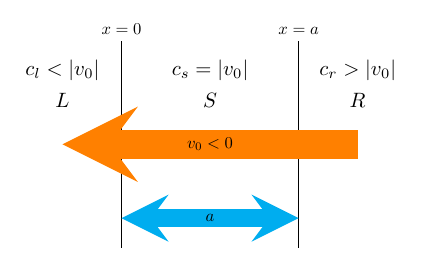} 
\caption{Acoustic black hole model: the homogeneous supersonic $L$ ($x<0$), sonic $S$ ($0<x<a$) and subsonic $R$ ($x>a$) regions are glued along the discontinuities at $x=0,a$.}
 \label{fsei}
\end{figure}

The model we shall study consists of a stepwise condensate (see Fig. (\ref{fsei})) formed by three homogeneous regions: $R$, i.e $x>a$, $S$, i.e $0<x<a$, $L$, i.e $x<0$.
Each region is characterized by its own constant external potential $V_{ext}^i$ %($i= R,S,L$)
 and constant coupling $g_i$ ($i= R,S,L$). The regions are glued along two step-like discontinuities at $x=a$ ($S-R$ junction) and $x=0$ ($L-S$ junction). We further assume that the condensate has everywhere uniform density $n$ as well as uniform flow velocity $v_0$ directed along the negative $x$ axis. The couplings $g_i$ are chosen so that the associated local speeds of sound $c_i=\sqrt{\frac{ng_i}{m}}$ satisfy the following requirements: in $R$  $c_r >|v_0|$, in $S$ $c_s=|v_0|$ and in $L$  $c_l<|v_0|$. So the flow is subsonic in R, supersonic in L and the transition occurs over an extended sonic region S of width $a$.

The values of $V_{ext}^i $ and $g_i$, albeit different in each region, have to satisfy the following constraint:
$V_{ext}^l+g_l n=V_{ext}^s+g_sn=V_{ext}^r+g_r n$.
Thanks to this condition the plane wave form Eq. (\ref{wfc}) of the condensate wave function $\Psi_0$ is solution of the Gross-Pitaevski Eq. (\ref{gp})
for all $x$ and time $t$.
Unlike the BH laser \cite{jac, refi},  in which a supersonic region is sandwiched between two subsonic regions, no instabilities appear here ($\omega$ is real).

%%%%%%%%%%%%%%%%%%%%%%%%%%%%%%%%%%%%%%%%%%%%%%%%%%%%%%%%%%%%%%%%%%%%%%%%%%%%%%%%%%%%%%%%%%%%%%%%%%%%%%%%%%%%%%%%%%%%%%%%%%%%%%%%%%%%%%%%%%%%%%%%%%%%%%
%A step-wise model of acoustic black hole was constructed by matching, along a discontinuity at $x=0$, a homogeneous supersonic ($x<0$) left region   %$L$ ($c=c_L>v=v_0$)  with a subsonic ($x>0$) right $R$ ($c=c_R>v=v_0$) one. It is displayed in Fig. (\ref{ftre}). The external potential $V_{ext}$ %and the repulsive atom-atom interaction coupling $g$
%are supposed to be constant in each region, but with different values in each region,  and satisfy
%$V_{ext}^l+g^l n=V_{ext}^r+g^r n$, guaranteeing that the same condensate wavefunction $\Psi_0$ is valid for all times $t$ and position $x$.

%\begin{figure}
%\includegraphics[width=0.8\columnwidth]{super-sub.jpg} 
%\caption{Two region model of acoustic black hole.
 %\label{ftre}}
%\end{figure}

\begin{figure}
\includegraphics[width=0.9\columnwidth]{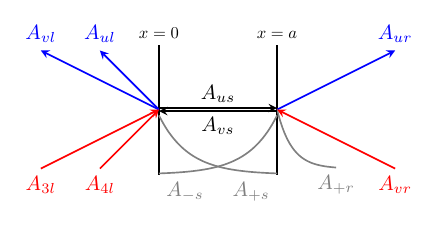} 
\caption{Matching between incoming modes (in red) and outgoing ones (in blue).}
%$4l$ `in mode': in red the incoming channel, in blue the outgoing ones
\label{fquattro}
\end{figure}

Full general solutions for this model  are obtained by imposing the general solution to obey the matching conditions at $x=0$ and at $x=a$
\begin{equation}\label{matchingaa}
[\phi]=0,\, [\phi']=0,\, [\varphi]=0,\, [\varphi']=0,
\end{equation}
with $[f(x)]=\lim_{\epsilon\to 0} [f(x+\epsilon)-f(x-\epsilon)]$ and $'$ means $\frac{d}{dx}$.
%to general solutions to the wave equations in the $L$ and $R$ regions
Note that because of stationarity of the model, $\omega$ is conserved.
Starting with the matching conditions at $x=0$ between the L and S regions we have that in L
 $\phi^L=\sum_{i} A_i\phi_\omega^i, i=ul,vl,3l,4l,$ and  in S $\phi^S=\sum_{j} A_j\phi_\omega^j, j=us,vs,+s,-s$ (and similarly for $\varphi$).  
We can write the 4 equations (\ref{matchingaa}) in matrix form 
\be
\label{emf}
M_lA_l=M_sA_s\ ,
\ee
 which relate the 4 left amplitudes $A_l\equiv (A_{ul},A_{vl}, A_{3l}, A_{4l}) $ to the 4 amplitudes in the sonic region $A_s\equiv (A_{us}, A_{vs}, A_{+s}, A_{-s})$.
We are considering here the case $\omega<\omega_{max}$ which is the interesting one. There is no Hawking-like emission for $\omega>\omega_{max}$.
The explicit form of the matrix $M_{l}$ is given by
\be
\label{eq:wl}
M_{l}=\left(
     \begin{array}{cccc}
       D_{vl} & D_{ul} & D_{3l} & D_{4l}\\
       ik_{vl}D_{vl} & ik_{ul}D_{ul} & ik_{3l}D_{3l} & ik_{4l}D_{4l} \\
       E_{vl} & E_{ul} & E_{3l} & E_{4l} \\
       ik_{vl}E_{vl} & ik_{ul}E_{ul} & ik_{3l}E_{3l} & ik_{4l}E_{4l}  \\
\end{array}\right)\,. 
\end{equation}
$M_s$ is similar in form up to the replacements $3,4\to +s,-s$.
%inserire eq. 1.80 del nostro " Understanding..."
%and $M_s$ can be constructed similarly.
We can proceed in the same way for the matching conditions at $x=a$ between the $S$ and the $R$ regions. Expressing the solution in $R$ as $\phi^R=\sum_{j} A_j\phi_\omega^j, j=ur,vr,+r,-r$ and similarly for $\varphi$ one ends with a matrix equation 
\be\label{ema}N_sA_s=N_rA_r\ , \ee 
where $A_r\equiv (A_{ur}, A_{vr}, A_{+r}, A_{-r})$. The matrices $N_{s(r)}$ have a form similar to (\ref{eq:wl}), up to the replacements $3,4\to +,-$ and the fact that now each entry is multiplied by the corresponding phase $e^{ik_{is(r)}a}$.  
We can combine (\ref{emf}) and (\ref{ema}) to eliminate the four amplitudes $A_s$ of the central sonic region and we get 
\be\label{me3}  M_lA_l=M_sN_{s}^{-1}N_r A_r\ .\ee
As standard in scattering theory, one proceeds with the construction of the ``in base" out of
`in' modes defined as those that start from $x=\pm\infty$ with amplitude $1$ and propagate towards the sonic region.  For $0<\omega<\omega_{max}$, at $x=-\infty$ they can be either characterized by $A_{3l}=1$ or $A_{4l}=1$ (the corresponding modes have  group velocity $\frac{d\omega}{dk}>0$) and, at $x=+\infty$, by $A_{vr}=1$ (the corresponding mode has $\frac{d\omega}{dk}<0$), see Fig. (\ref{fquattro}). 
The channel that leads to Hawking radiation corresponds to
%For the particular case of the `in' $4l$ mode, depicted in Fig. (\ref{fquattro}), we choose 
$A_{4l}=1,  A_{3l}=0, A_{vr}=0, A_{-r}=0$ (the amplitude $A_{-r}$ of the growing mode, which explodes at $x=+\infty$, is always set to $0$ to ensure normalizability) 
%. This is the channel which leads to Hawking like radiation 
\cite{capitolo-libro}. 
The 4 matching equations (\ref{me3}) will explictly determine the remaining amplitudes $A_{vl},A_{ul}, A_{ur}, A_{-r}$. 
%Being the $4l$ mode of %negative norm the amplitudes appearing in Fig. (\ref{fquattro}) satisfy the unitarity condition 
%$|A_u^r|^2+|A_v^l |^2-|A_u^l |^2=-1$. 
Similar constructions apply for the positive norm `in' modes  $3l$ and $vr$. 

%The three 'in' modes altogether constitute a basis (the 'in' basis) for the fluctuation field $\hat\phi$.  
An alternative basis, the `out' basis, is formed by three `out' modes, i.e. those that propagate away from the sonic region towards $x=\pm\infty$. The ones that propagate towards $x=-\infty$ are either $A_{vl}=1$ (positive norm) or $A_{ul}=1$ (negative norm), both with negative group velocity. For $x=+\infty$ we have $A_{ur}=1$ (positive norm and positive group velocity). 

%They are named `out' $ul$ (negative norm), $vl,ur$ (positive norm) modes. 

We can expand the fluctuation field $\hat\phi$ (\ref{frep}) in the `in' and `out' basis with the corresponding annihilation and creation operators $\hat a_\omega^{i, in(out)}, \hat a_\omega^{i, in(out)\dagger}$ (for the `in' basis $i=4l,3l,vr$ and for the `out' basis $i=vl,ul,ur$). 
The presence of negative norm modes implies that the `in' and `out' vacuum states (i.e. the states annihilated by all the $\hat a_\omega^{i,in(out)}$)  are not the same. This implies the spontaneous creation of phonons. In particular,   
by preparing $\hat\phi$ in the `in' vacuum state the number of spontaneously created $k_{ur}$ phonons (the analogous of Hawking BH radiation) per unit time and per unit bandwidth in region R is given by 
\be \label{nur}n_\omega^{ur}=|A_{ur}|^2\ .\ee

\begin{figure}
\includegraphics[width=1\columnwidth]{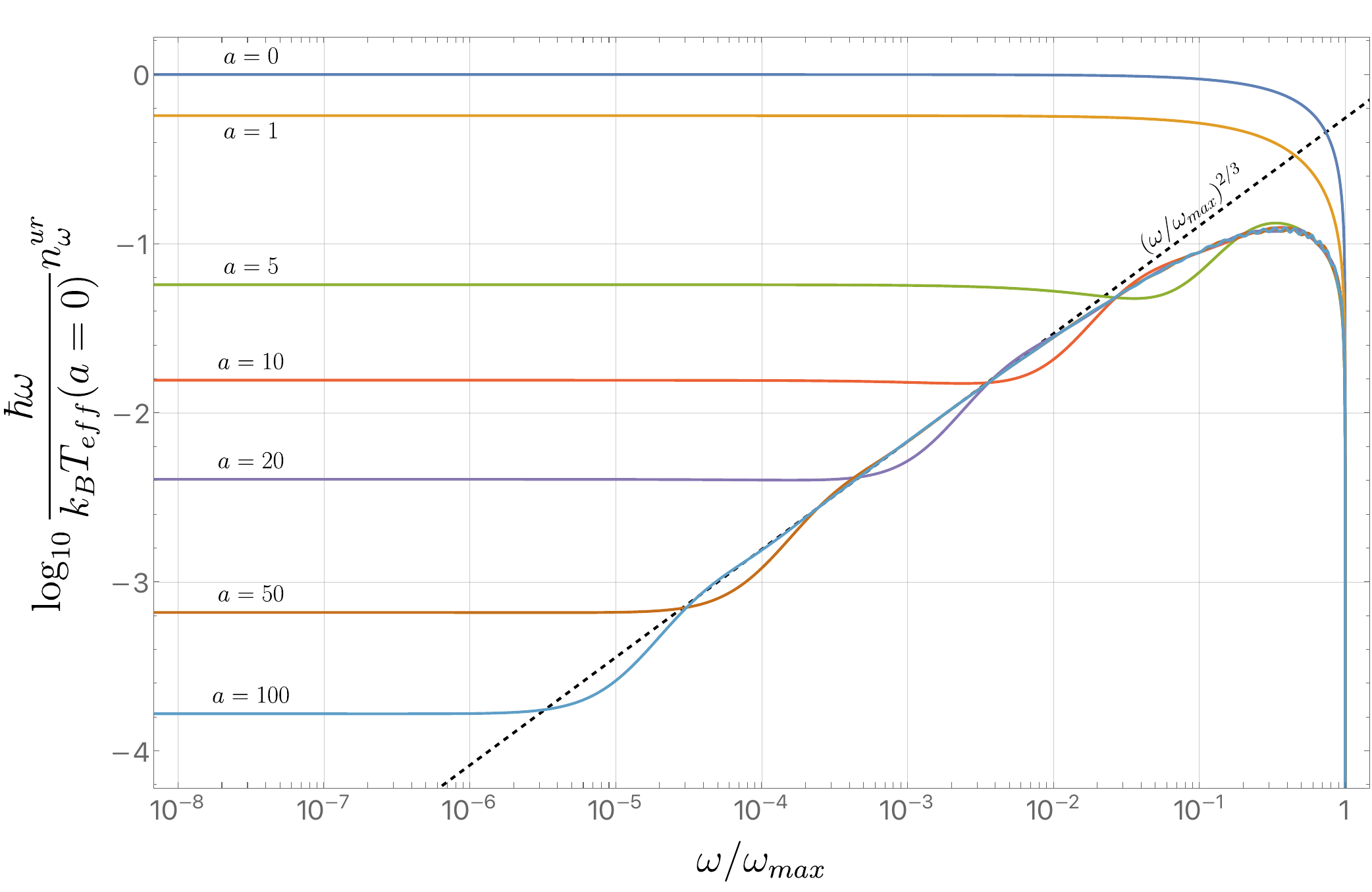} 
\caption{Log-log plot of $\frac{\hbar\omega}{k_B T_{eff}(0)} n_\omega^{ur}$ for various values of $a$.
Here and in the figures that follow $c_l=0.5, c_s=|v_0|=1, c_r=2$ ($\hbar=1,\ m=1$). From the initial $\omega n_\omega^{ur} \sim const$ behaviour we see the transition to the regime where $\omega n_\omega^{ur} \sim \omega^{2/3}$ (i.e. $n_\omega^{ur}\sim \omega^{-1/3}$).
\label{fdieci}}
\end{figure}

\begin{figure}
\includegraphics[width=1\columnwidth]{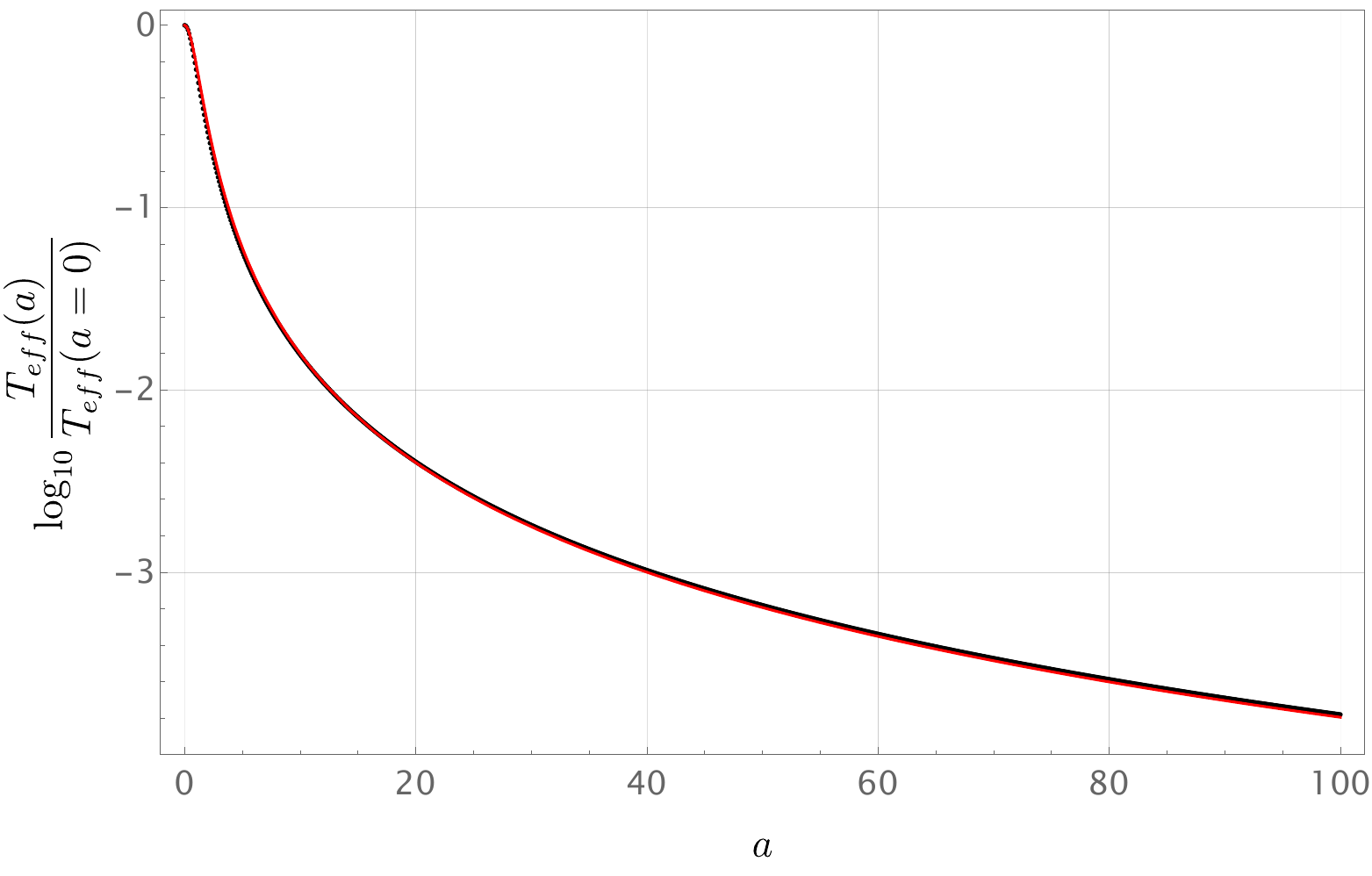} 
\caption{Log plot of $\frac{T_{eff}(a)}{T_{eff}(0)}$. In black is the numerical curve, in red the fitted curve $\frac{1}{Aa^2+1},\ A=0.646$.
\label{fundici}}
\end{figure}

When there is no intermediate sonic region (i.e. $a=0$) it has been shown \cite{rpc, mfr, lrpc, capitolo-libro} that, for small $\omega$, $n^{ur}_\omega$ has a thermal $\frac{1}{\omega}$ behavior namely 
 $n^{ur}_\omega \sim \frac{ k_B T_{eff}(a=0)}{\hbar \omega}$
where \cite{capitolo-libro} %$T(0) = h(c_l-v_0).....
\begin{equation}
\mathrm{T_{eff}(a=0)}=\frac{\hbar}{k_B} \frac{(c_r+v_0)}{(c_r-v_0)}\frac{(v_0^2-c_l^2)^{3/2}}{(c_r^2-c_l^2)}\frac{2c_r}{c_l\xi_l}
\end{equation}
can be identified as a sort of effective temperature.
A similar pattern is found even in the presence of a sonic region ($a\neq 0$). Looking at Fig (\ref{fdieci}) we see clearly the $\frac{1}{\omega}$ behavior at small $\omega$ but with an effective temperature $T_{eff}(a)$  (which can be read from the intersection of the the various curves with the vertical axis) that rapidly decreases as the width $a$ of the sonic region increases (see Fig. (\ref{fundici})) .This trend is similar to the one found in Refs. \cite{iacmax,  zapa} where a resonant cavity was inserted in the supersonic region. A numerical fit gives in our case 
$\frac{T_{eff}(a)}{T_{eff}(0)} \sim \frac{1}{0.646a^2+1}$. %\ A=0.646$.  
This expression %on the r.h.s. 
can be considered as a sort of (low frequency) gray-body factor caused by the backscattering of the modes in the sonic cavity. Its independence on $\omega$  is consistent with the expectations for acoustic BHs (see \cite{rigorous}) and should be compared for example to the Schwarzschild black hole where instead it scales as $\omega^2$ \cite{page76}.

Inspection of Fig(\ref{fdieci}) further reveals that the extension of the thermal $\frac{1}{\omega}$ region decreases as $a$ increases and a transition to a $\frac{1}{\omega^{1/3}}$ behavior occurs. Note that this scaling would be the leading one on a small $\omega$ expansion if there were no subsonic region (i.e. $a\to  \infty$) \cite{in-prep}. Thus in this case the emission is not thermal.
One further notices from Fig. (\ref{fdieci}) that while for $\omega\to 0$  $n^{ur}_\omega$ decreases rapidly by orders of magnitude when increasing the width of the sonic region, when $\omega$ is of order $10^{-1}$ the signal is independent on $a$ when $a \gtrsim 5$.

\begin{figure}
\includegraphics[width=1\columnwidth]{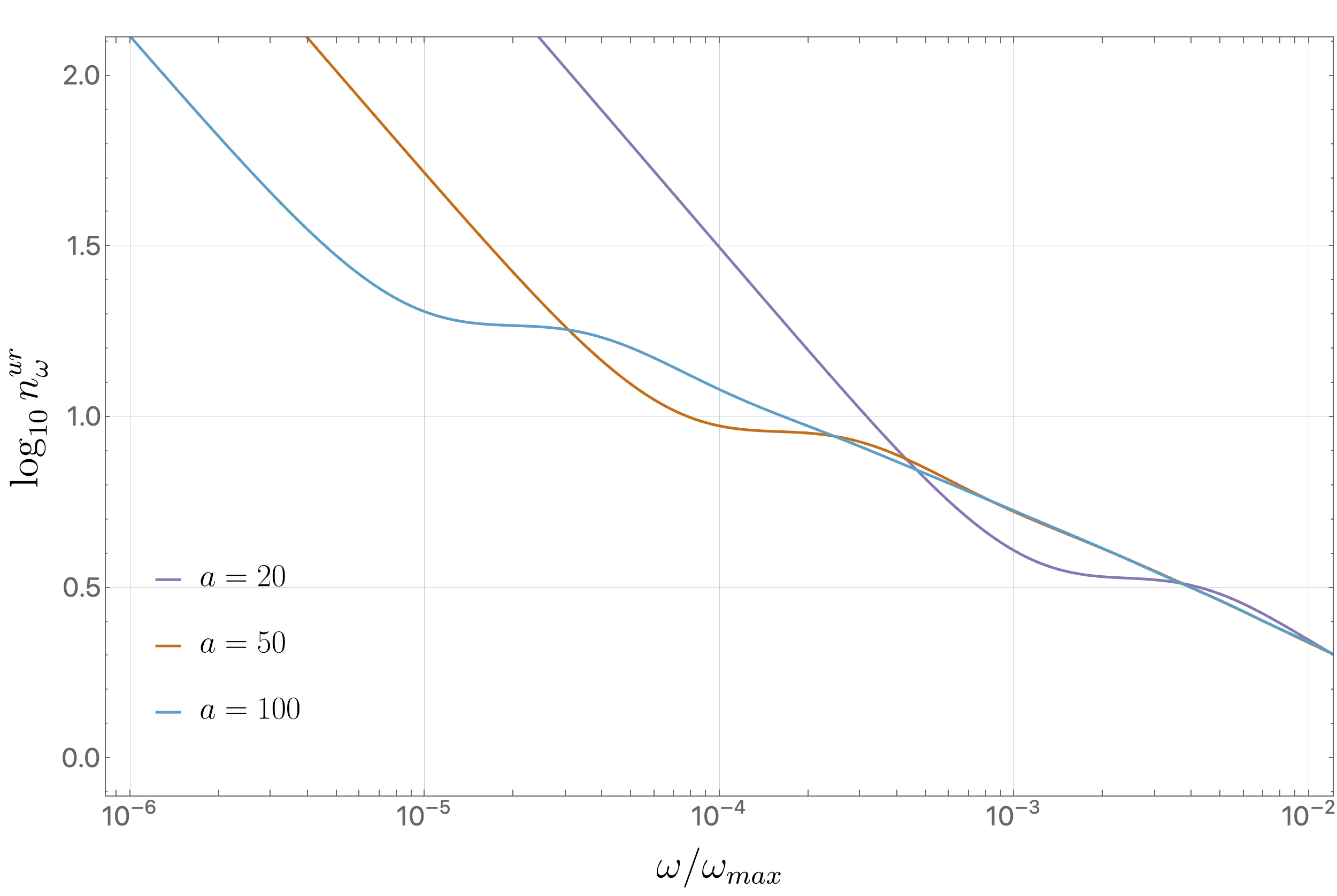} 
\caption{Log-log plot of $n_\omega^{ur}$ at the onset of the oscillations for $a=20,50,100$.
 \label{fdodici}}
\end{figure}

\begin{figure}
\includegraphics[width=1\columnwidth]{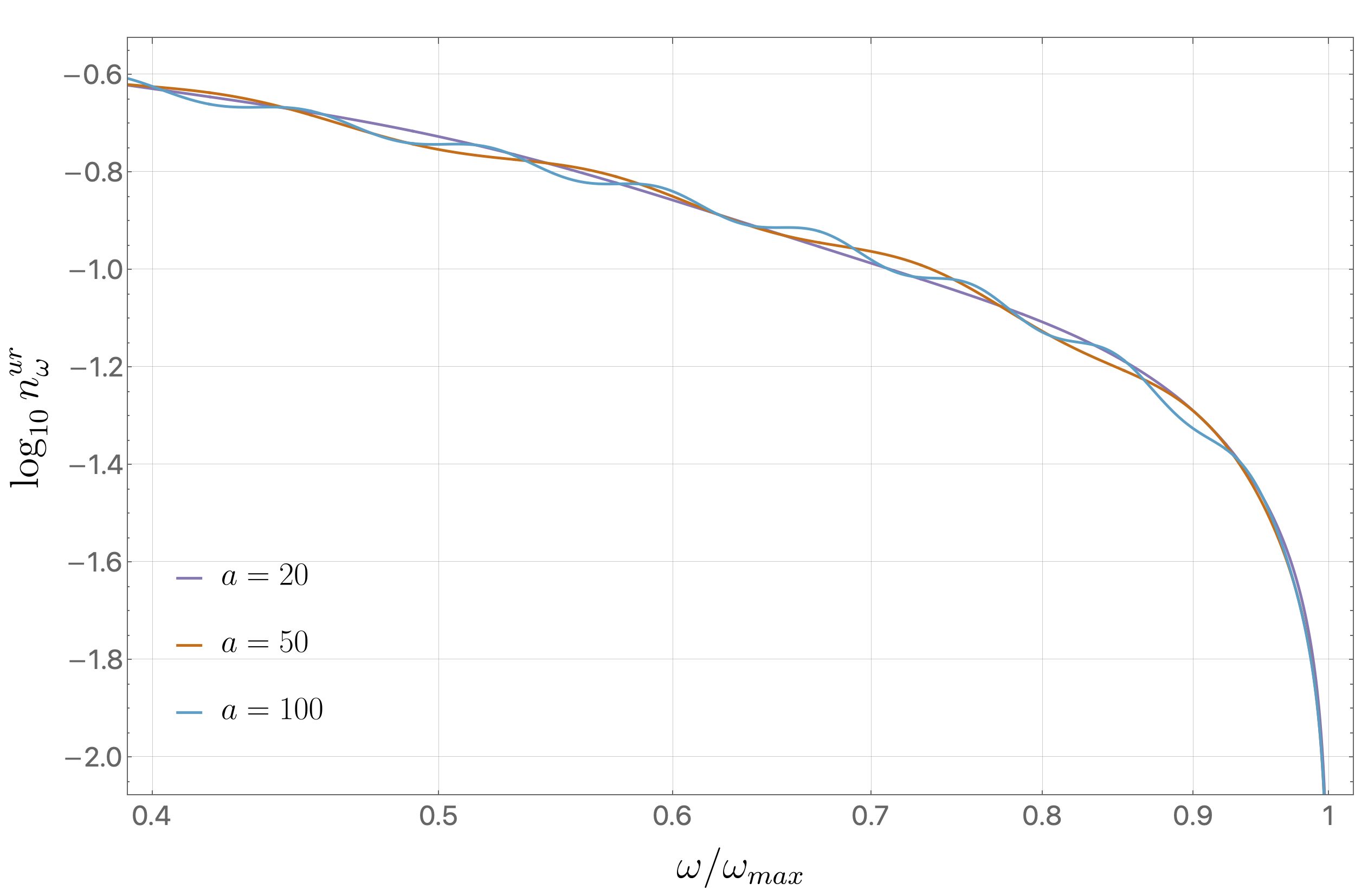} 
\caption{Log plot of $n_\omega^{ur}$ for $a=20,50,100$ as $\omega_{max}$ is approached.
 \label{ftredici}}
\end{figure}

Interestingly,  the $\frac{1}{\omega^{1/3}}$ phase is characterized by small oscillations, a sort of modulation of the signal which can be seen in Figs (\ref{fdodici}, \ref{ftredici}) where a magnification of the scale is performed 
and a limited interval of $\omega$  is considered to better appreciate this feature. 
In Fig. (\ref{fdodici}) we have the onset of the oscillations and in Fig. (\ref{ftredici}) the damping of $n_\omega^{ur}$ as $\omega_{max}$ is approached. In Fig. (\ref{ftredici}) one can also see that the period (in $\omega$) of the oscillations varies with $a$: as $a$ increases the period decreases.

In conclusion we have seen how modes propagate in a purely sonic region and that ``broadening the horizon'' drastically reduces the Hawking signal, as inferred in \cite{fipa}, and induces a modulation of the signal that depends on its width. This behavior is quite different from the one emerging when a resonant cavity is placed in a supersonic region. In this case isolated sharp resonances appear \cite{zapa, iacmax}.

\textbf{Data availability:} The data supporting the findings of this paper are openly available \cite{daniel}.

\textbf{Acknowledgments:} We thank N. Pavloff and S. Robertson for useful discussions. 
A.F. and D.P. acknowledge partial financial support by the Spanish Grants PID2020-116567GB-C21, PID2023-149560NB-C21 funded by MCIN/AEI/10.13039/501100011033 and FEDER, European Union, and the Severo Ochoa Excellence Grant CEX2023-001292-S. D.P. acknowledges the `Atracci\'o de Talent' program of the University of Valencia for a doctoral grant.

\end{document}